\documentclass[twocolumn,aps,prl,showpacs]{revtex4}
\usepackage{graphicx}
\usepackage{natbib}
\usepackage{amsbsy,amssymb,amsmath}
\usepackage{dcolumn}
\usepackage[usenames]{color}
\usepackage{psfrag}
\usepackage[colorlinks,bookmarks]{hyperref}
\definecolor{linkblue}{rgb}{0,0,0.8}
\definecolor{linkgreen}{rgb}{0,0.5,0}
\hypersetup{pdfpagemode=UseNone, pdfstartview=FitH, linkcolor=linkblue, %
            citecolor=linkgreen, urlcolor=linkblue}

\newcommand{\beq}{\begin{equation}}
\newcommand{\eeq}{\end{equation}}

\def\lcdm{$\Lambda$CDM}

\def\mb{M_{\rm b}}
\def\vf{V_{\rm f}}
\def\vvir{V_{\rm vir}}

\begin{document}

\title{What do gas-rich galaxies actually tell us about modified Newtonian dynamics?}

\author{Simon Foreman}
\email{sfore@stanford.edu}
\altaffiliation[Current address: ]{Department of Physics, Stanford University, Stanford, CA 94305 USA}
\affiliation{Department of Physics and Astronomy, %
University of British Columbia, %
Vancouver, BC, V6T 1Z1  Canada}

\author{Douglas Scott} 
\email{dscott@phas.ubc.ca}
\affiliation{Department of Physics and Astronomy, %
University of British Columbia, %
Vancouver, BC, V6T 1Z1  Canada}

\date{\today}

\begin{abstract}
It has recently been claimed that measurements of the baryonic Tully-Fisher relation (BTFR), a power-law relationship between the observed baryonic masses and outer rotation velocities of galaxies, support the predictions of modified Newtonian dynamics for the slope and scatter in the relation, while challenging the cold dark matter (CDM) paradigm. We investigate these claims, and find that: 1) the scatter in the data used to determine the BTFR is in conflict with observational uncertainties on the data; 2) these data do not make strong distinctions regarding the best-fit BTFR parameters; 3) the literature contains a wide variety of measurements of the BTFR, many of which are discrepant with the recent results; and 4) the claimed CDM ``prediction" for the BTFR is a gross oversimplification of the complex galaxy-scale physics involved. We conclude that the BTFR is currently untrustworthy as a test of CDM.
\end{abstract}


\maketitle

{\em Introduction.---}%
Despite the plentiful evidence for the existence of dark matter in the Universe (see Ref.~\cite{dm-review} for a recent review), the modified Netwonian dynamics (MOND) hypothesis---that Newtonian gravity departs from its expected behaviour below a certain acceleration scale, thus potentially eliminating the need for any non-luminous, non-baryonic matter~\cite{milgrom-original}---has persisted since its proposal almost three decades ago. It has recently been claimed by McGaugh in Ref.~\cite{mcgaugh2011} (hereafter MG11) and in a follow-up paper~\cite{mcgaugh-apj2011} that measurements of the baryonic Tully-Fisher relation (BTFR)~\cite{mcgaugh-btfr-2000,freeman-btfr}, a scaling relation between the baryonic masses and outer rotation velocities of galaxies, match precisely with a prediction of MOND involving {\em zero} free parameters, with scatter about this relation attributable solely to observational uncertainties. MG11 also claims that the observed BTFR deviates significantly from that predicted by the standard cold dark matter (CDM) scenario. Investigations into each of these claims reveal flaws that substantially weaken the conclusions of MG11, and highlight the difficulty of using such an apparently simple relationship as the BTFR as a clean test of new physics.

The data for the 47 galaxies used in MG11 are collected from three sources \cite{stark2009,begum2008,trachternach2009}, subject to the criteria that the mass of molecular gas in each galaxy exceeds its stellar mass, and that each galaxy has a resolved rotation curve that asymptotes to a constant velocity at large radius. We have assembled our own sample of galaxies (henceforth FS11) from the same three sources, subject to the same two criteria and re-calculating derived quantities where needed.

We find 58 galaxies that are suitable for use. In brief, we include 11 galaxies from Ref.~\cite{stark2009} and one from Ref.~\cite{begum2008} that we find to meet the necessary criteria but that are excluded from MG11, and exclude one galaxy used by MG11 that we find to have star-dominated mass. Many of the extra galaxies have low inclinations ($i<45^\circ$), which have been seen to decrease the slope of the BTFR~\cite{stark2009}, but we see no clear quality issues that should cause these data should be excluded (especially since MG11 utilizes galaxies with inclinations as low as~$29^\circ$).

However, we should note that the criteria of MG11 likely impart strong selection effects and biasing in the determination of the BTFR. For example, as discussed by Ref.~\cite{gurovich2010}, the observability of a flat region in a rotation curve favours galaxies for which the distribution of neutral hydrogen is more extended compared to the radial scale of the outer halo---indeed, many otherwise well-behaved galaxies do not possess flat rotation curves~\cite{salucci-rc}. In addition, galaxies undergoing significant interaction with others nearby will tend to be disfavoured. Further criteria based on data quality are sometimes used in BTFR studies~\cite{mcgaugh-apj2011}, but imposing such criteria can tend to skew the sample towards galaxies with particular properties, when in fact the goal should be to use a broad range of data to avoid selection affects biasing the results.

{\em Scatter in galaxy data.---}%
The strategy of MOND is most simply characterized by its suggestion that the acceleration $\vec{a}_{\rm N}$ of a test particle in Newtonian gravity is actually related to the observed acceleration $\vec{a}$ by the relation
\beq
\label{eq:mondeq}
\vec{a}_{\rm N} = \mu(a/a_0) \vec{a},
\eeq
where $a_0$ is a new physical constant with dimensions of acceleration and magnitude roughly $10^{-10}$ ${\rm m} \, {\rm s}^{-2}$, and $\mu(x)$ is a smooth function that asymptotes to unity for $x\gg1$ and to $x$ for $x\ll1$. The modern perspective is that the relation (\ref{eq:mondeq}) might arise from some fully covariant theory of non-Einsteinian gravity \cite{bekenstein2004,milgrom-bimond,clifton-mgreview}, although such theories continue to face difficulties of their own \cite{skordis-teves,bruneton-covmond,ferreras-teves,contaldi-teves}.

Upon taking the acceleration of a test particle within a galaxy to satisfy $a \ll a_0$, and assuming that galaxies contain only baryonic matter, simple algebra yields the MOND prediction for the BTFR:
\beq
\label{eq:mondbtfr}
\mb = \frac{\vf^4}{Ga_0},
\eeq
where $\mb$ is the (baryonic) mass of a galaxy and $\vf$ is the rotation speed asymptotically approached at large radius. For disk galaxies, $a_0$ should be multiplied by an extra factor $\chi\approx0.8$ to account for the difference in rotation speeds between disk-like and spherical mass distributions~\cite{mcgaugh-apj2011}. A good indicator of the scatter of a sample of galaxies about this relation is the distribution of $a_0$ values derived from the measured values of $\mb$ and $\vf$ for the sample.

\begin{figure}
\centering \mbox{\resizebox{0.48\textwidth}{!}{\includegraphics[angle=0,trim=0 0 0 0]{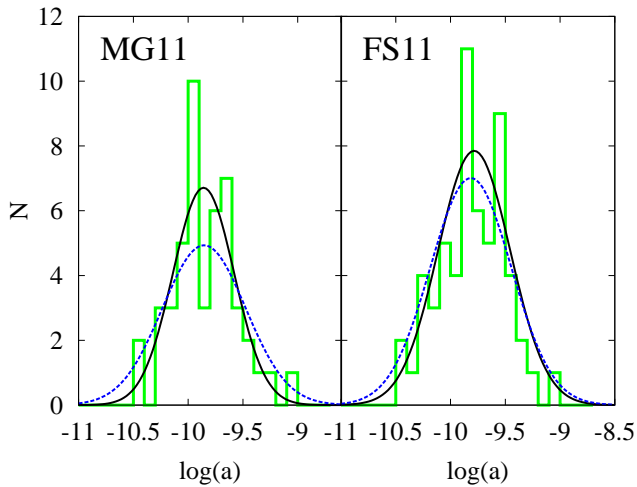}}}
\caption{Histograms (green bars) of values of $a=\vf^4/G\mb$ (in units of ${\rm m} \, {\rm s}^{-2}$) for MG11 and FS11 data sets, along with fitted Gaussians (black, solid curves) and Gaussians with the most likely mean and variance for data sets with the same statistical properties as the measured data (blue, dashed curves). For MG11, the Monte Carlo-generated distribution is much wider than the fitted distribution, indicating a discrepancy in the level of scatter in the data as compared to the stated error bars. This discrepancy is less severe for FS11. \label{fig:myhists}}
\end{figure}

Therefore, analogously to MG11's figure 3, we calculate $a\equiv\vf^4/G\mb$ for each of the 47 galaxies used by MG11. In Fig.~\ref{fig:myhists},  we display a histogram of these values, as well as a Gaussian fit to this histogram. The fitted Gaussian has width $\sigma=0.28$ dex, similar to the width of the Gaussian from MG11 (which was not fitted, but rather estimated by eye), quoted as $0.24$ dex.

To put the scatter in $a$ values in perspective, we generate a large number ($10^7$) of simulated datasets of 47 galaxies [i.e.\ $(\vf, \mb)$ pairs], with the velocity and baryonic mass of each galaxy drawn from a normal distribution determined by the error bars on the corresponding galaxy from the MG11 sample. We calculate the mean and variance of the $a$ values for each dataset, and from this determine the most likely mean and variance of a galaxy sample with the same properties as that of MG11.  We plot a Gaussian with the resulting mean (${\rm log} [a_0] = -9.85$) and variance ($\sigma=0.38$ dex) as the blue dashed line in Fig.~\ref{fig:myhists}. The width of our Monte Carlo-generated distribution is much larger than that of the data values themselves (regardless of whether the latter width is taken to be $0.24$ or $0.28$ dex), so the data somehow show {\em less} scatter than would be expected from the errorbars. MG11's claim that the scatter in the data is explained by observational uncertainty alone overlooks this statistical feature.

Performing the same analysis on the FS11 data, we find that a Gaussian fitted to the calculated $a$ values has $\sigma=0.33$ dex. Using a Monte Carlo procedure as above to determine the most likely Gaussian that fits the data with the given errorbars, we find ${\rm log} (a_0) = -9.82$ and $\sigma=0.37$ dex. The FS11 data therefore show increased scatter as compared with the MG11 data, and although the scatter is still not completely accounted for by observational uncertainty, there is less of a discrepancy than for the MG11 sample.

{\em Fitting the BTFR.---}%
We determine the values of the BTFR parameters ($a_0$ and the slope, $\alpha$) by minimizing the $\chi^2$ statistic that accounts for errors in both ${\rm log} (\mb)$ and $\vf$, given in this case %
\footnote{The first version of this paper contained a sign error in Eq.~(\ref{eq:chi2}), but all calculations used the correct formula, so the original numerical results were not affected.} %
by 
\beq
\label{eq:chi2}
\chi^2 = \sum_i \frac{ \left[ {\rm log} (\mb) - \alpha \, {\rm log} (\vf) + {\rm log} (Ga_0) \right]^2}
	{\sigma_{{\rm log} (\mb)}^2 + \alpha^2 \sigma_{{\rm log} (\vf)}^2},
\eeq
where $\sigma_{{\rm log} (\vf)}$ is determined by the asymmetric error bars on ${\rm log} (\vf)$ that follow from assuming Gaussian uncertainties for $\vf$. This expression does not account for intrinsic scatter, since MOND predicts that it should be precisely zero.

\begin{figure}
\centering \mbox{\resizebox{0.485\textwidth}{!}{\includegraphics[angle=0,trim=0 0 0 0]{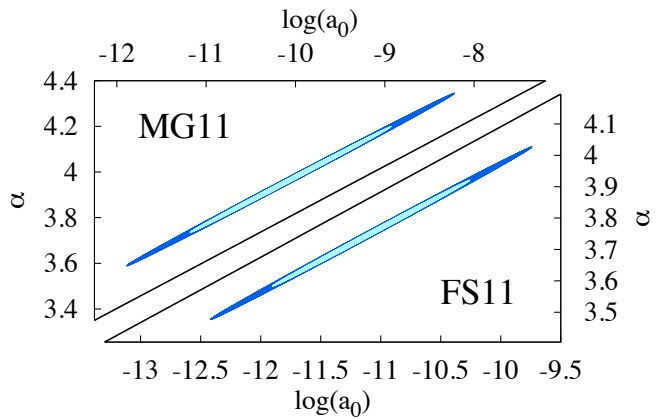}}}
\caption{68\% (light blue, inner) and 95\% (dark blue, outer) confidence regions for the best-fit values of $a_0$ (in units of ${\rm m} \, {\rm s}^{-2}$) and $\alpha$ for the MG11 and FS11 data sets (plotted on different sets of axes). It is apparent that the quality of the data makes it difficult to distinguish $\alpha=4$ (the prediction of MOND) from lower values; this difficulty is reflected in the wide range of slopes measured in other studies~
\cite{stark2009,trachternach2009,meyer2008,belldejong2001,noordermeer2007,geha2006,
torresflores2011,pfenniger2005,kregel2005,
kassin2007,derijcke2007,avilareese2008,gurovich2010,
verheijen2001,mcgaugh-apj2005}.
\label{fig:ells}}
\end{figure}

\begin{table*}
\begin{center}
\setlength{\extrarowheight}{1.5pt}
\begin{tabular}{|c||c|c|c|c|c|c|}
\hline
	& \multicolumn{6}{c|}{\textbf{best-fit parameters}} \\
	\hline
	& \multicolumn{2}{c|}{\textbf{$\boldsymbol{a_0}$ free, $\boldsymbol{\alpha=4}$ fixed}}
	& \multicolumn{4}{c|}{\textbf{both $\boldsymbol{a_0}$ and $\boldsymbol{\alpha}$ free}} \\
	\hline
	\multicolumn{1}{|c||}{{\textbf{data}}} & 
	\multicolumn{1}{c|}{{$a_0 \times 10^{10}$ (${\rm m} \, {\rm s}^{-2}$)}}  &
	\multicolumn{1}{c|}{{$\chi^2$}} &
	\multicolumn{1}{c|}{{$a_0 \times 10^{10}$ (${\rm m} \, {\rm s}^{-2}$)}}  &
	\multicolumn{1}{c|}{{${\rm log} (a_0 / [ {\rm m} \, {\rm s}^{-2} ])$}} & \multicolumn{1}{c|}{{$\alpha$}} &
	\multicolumn{1}{c|}{{$\chi^2$}} \\
	\hline
      MG11 (47 galaxies)	& $1.27 \pm 0.09$ & 65.0 & $0.76^{+10.24}_{-0.70}$
      									& $-10.1 \pm 1.1$ & $3.96 \pm 0.23$ & 64.9  \\
      FS11 (58 galaxies)	& $1.33 \pm 0.08$ & 137.8 & $0.08^{+0.47}_{-0.07}$
      									& $-11.1 \pm 0.8$ & $3.75 \pm 0.17$  & 133.0 \\
      \hline
\end{tabular}
\caption{\label{tab:c2fit} BTFR parameters that minimize the value of $\chi^2$ [Eq.~(\ref{eq:chi2})] for the MG11 and FS11 data sets. The effect of the disk rotation factor described below Eq.~(\ref{eq:mondeq}) has been included in the calculation of $a_0$. When $\alpha$ is left free, it is more natural to state the constraints on $a_0$ in terms of its logarithm, but we also give confidence intervals for $a_0$ itself for comparison with the fixed-slope case. The lowest $\chi^2$ values we find are much larger than those found in MG11 (around~44).}%
\end{center}
\end{table*}

The parameter values that minimize Eq.~(\ref{eq:chi2}) are given in Table~\ref{tab:c2fit}. The uncertainties on these values were determined by the limits of the 68\% confidence regions defined by $\chi^2 < \chi^2_{\rm min} + \Delta\chi^2$, where $\Delta\chi^2=1.0$ for a one-parameter fit or 2.3 for a two-parameter fit (see, e.g., Ref.~\cite{rpp}). Observe that the lowest $\chi^2$ value we can obtain, 64.9, is far from the $\sim$44 stated by MG11.

Examining the confidence regions for the two-parameter fits (shown in Fig.~\ref{fig:ells}), we find a broad range of slopes that can provide acceptable agreement with the data. A value of $\alpha=4$ is contained within the 95\% confidence regions for both data sets, but these regions also cover slopes down to 3.6 for MG11 and 3.5 for FS11, a large range that reflects the variety of slopes measured in other studies (which we describe below). Either data set can only weakly discriminate between the MONDian slope ($\alpha=4$) and one bearing no connection to MOND ($\alpha\neq 4$).

{\em Other measurements of the BTFR.---}%
There are several studies that provide estimations of the BTFR slope from different galaxy samples, utilizing a wide variety of techniques for measuring $\vf$ and $\mb$ for each galaxy. Some find that slopes $\sim$4 best describe their data \cite{mcgaugh2011,mcgaugh-apj2011,stark2009,trachternach2009,meyer2008}, while others find values roughly midway between 3 and 4 \cite{belldejong2001,noordermeer2007,geha2006,torresflores2011,pfenniger2005,kregel2005} or lower still \cite{kassin2007,derijcke2007,avilareese2008,gurovich2010}, or alternately find a sensitive dependence on the methods used to convert luminosity to stellar mass \cite{verheijen2001,mcgaugh-apj2005}. Useful summaries of these measurements are contained in Refs.~\cite{gurovich2010} and~\cite{avilareese2008}. Many of these studies stand in contrast with the results of MG11. Also, there are indications that if the mass of neutral hydrogen in a galaxy is used instead of baryonic mass, the resulting Tully-Fisher relation strongly favours Newtonian gravity over MOND~\cite{chakraborti-h1}.

Various arguments exist as to why some of the above measurements might be flawed. For example, Ref.~\cite{mcgaugh-apj2011} claims that the data used by Ref.~\cite{gurovich2010} are not normally-distributed and hence cannot be used in a least-squares fit, and also that the rotation curves of Ref.~\cite{torresflores2011} might not be extended enough to supply accurate values for $\vf$. Meanwhile, Ref.~\cite{gurovich2010}, conjectures that MG11's selection criteria of flat rotation curves could bias the determination of $\alpha$ toward higher values, and Ref.~\cite{torresflores2011} highlights issues surrounding the assembly of galaxy samples from several different observational surveys.

When possible sample selection biases and differences in observational and data analysis techniques are taken into account, it becomes extremely difficult to rank the above studies in terms of accuracy of their estimations of the BTFR. It is clear, however, that the question of ``the" slope of the relation is far from settled, certainly not at the level needed to test a theory predicting a value between 3 and 4.

{\em Cold dark matter and the BTFR.---}%
Figure 2 of MG11 plots that paper's galaxy data along with the MOND-predicted BTFR, Eq.~(\ref{eq:mondbtfr}), and a line taken to represent the BTFR anticipated in the CDM scenario. What is used for the latter is a cosmologically-motivated scaling relation (see, e.g., Ref.~\cite{white-scaling}), which has
$\mb \propto \vf^3$,
under the assumptions that, for each galaxy, $\mb$ is equal to the cosmic baryon fraction times the virial mass  and $\vf$ is equal to the rotation velocity at the virial radius. Neither of these assumptions is well-motivated, since the baryon fraction has been seen to vary from galaxy to galaxy (Ref.~\cite{torresflores2011} provides a recent example of this), and, as mentioned by Ref.~\cite{gurovich2010}, the theory of Navarro-Frenk-White profiles leads us to believe that $\vf$ is not representative of $\vvir$. Ref.~\cite{mcgaugh-apj2011} takes steps towards improving these assumptions by introducing extra fitting factors into the relations between the virial and observed quantities,  but still overlooks a fundamental problem with this approach.

The essence of the problem is that galaxies are complex objects, with individual histories and properties determined by feedback mechanisms, ``gastrophysics," and a whole host of processes that are not yet fully understood. Therefore, a first-principles calculation of an analytical relationship between masses and rotation velocities of galaxies that is expected to hold in general is simply not possible in the context of CDM---a complete numerical calculation of the detailed physics that would affect such a relationship is still beyond our reach. Semianalytic treatments (e.g.~\cite{bullock2001}) can make some progress, but in reality, these efforts tend to be tuned to match observations of the Tully-Fisher relation, rather than providing predictions. In this light, the fact that the supposed prediction of {\lcdm} is nowhere near the data in MG11's figure 2 is misleading---there are certainly published estimates for the BTFR, obtained through techniques like halo abundance matching~\cite{trujillo-ham}, as well as models for feedback processes~\cite{piontek-feedback,derossi-feedback}, that show tentative agreement with current data. Ref.~\cite{mcgaugh-apj2011} describes how these studies are imperfect in various ways, but this is only to be expected, as our understanding of galactic physics is not yet complete.

{\em Discussion.---}%
MG11's conclusion, that the scatter-free BTFR describing a certain sample of galaxies strongly favours MOND over {\lcdm}, is a bold one, and should be evaluated carefully and cautiously. We have raised concerns about the scatter of the data sample, the data's preference (or lack thereof) for a certain BTFR slope, the wide variety of measured slopes found in the literature, and what MG11 claims the theory of CDM has to say about the BTFR. These concerns are supplemented by those of Ref.~\cite{gnedin2011}, which highlights how the measured baryonic masses in MG11 do not include the ionized gas content of galaxies; models that take this into account could potentially provide reasonable agreement both with the observed data and with recent N-body simulations of dark matter halos~\cite{prada-halos}.

MOND, treated as a phenomenological description of galaxy-scale physics (particularly, galaxy rotation curves~\cite{sanders-rotcurves,swaters-rotcurves} and the properties of dwarf galaxies~\cite{kosowsky-monddwarfs}), is apparently quite successful. However, the tenets of MOND are challenged in the same regime by models of the globular cluster NGC~2419 \cite{ibata-crucible,ibata-polytrope} and fits to the ``neutral hydrogen Tully-Fisher relation" \cite{chakraborti-h1}. As well, the scalings and value of $a_0$ are unsurprising when put in the context of familiar facts about galaxies and cosmology \cite{scott-findumond}.

Due to the ambiguity present in current data and in our picture of the behaviour of galaxies, the BTFR provides no new advantage for MOND over CDM. Nevertheless, as the data improve and systematics become less significant, the BTFR could begin to provide useful information. In particular, more precise measurements of $\alpha$ could serve as a guideline with which new models for supernova feedback would have to agree, and these measurements could therefore assist in the development of a full description of the baryonic processes that dominate deep within CDM haloes. It is also worth exploring variants of the BTFR, using other mass tracers like neutral hydrogen \cite{chakraborti-h1} or considering the relation at different radii \cite{yegorova-rtf} (although it is nontrivial to define a suitable radius consistently for gas-dominated galaxies). If supplemented with other observational probes, the BTFR could well provide a unique window into the realm of galactic physics.


{\em Acknowledgements.---}%
This research was supported by the Natural Sciences and Engineering Research Council of Canada. We thank Paolo Salucci for useful comments.

\bibliography{mond_bib}

\end{document}